\documentclass[11pt,nofootinbib,showpacs]{revtex4}
\usepackage{amsfonts,amssymb,amsmath,graphicx,multirow,dsfont}
\def\dac{\displaystyle\frac}

\def\[{\left[}
\def\]{\right]}
\def\({\left(}
\def\){\right)}

\def\gammad{\gamma _{\left( \mathbf{D}\right)}}
\def\gammat{\gamma _{\left(3 \right)}}

\newcommand{\diag}{\mathop{\rm diag}\nolimits}

\begin{document}

\baselineskip7mm

\title{Effects of spatial curvature and anisotropy on the asymptotic regimes in Einstein-Gauss-Bonnet gravity}

\author{Sergey A. Pavluchenko}
\affiliation{Programa de P\'os-Gradua\c{c}\~ao em F\'isica, Universidade Federal do Maranh\~ao (UFMA), 65085-580, S\~ao Lu\'is, Maranh\~ao, Brazil}
\author{Alexey Toporensky}
\affiliation{Sternberg Astronomical Institute, Moscow State University, Moscow 119991 Russia}
\affiliation{Kazan Federal University, Kazan 420008 Russia}

\begin{abstract}
In this paper we address two important issues which could affect reaching the exponential and Kasner asymptotes in Einstein-Gauss-Bonnet cosmologies -- spatial curvature and anisotropy in
both three- and extra-dimensional subspaces. In the first part of the paper we consider cosmological evolution of spaces being the product of two isotropic and spatially curved subspaces.  It is demonstrated that the dynamics in $D=2$ (the number of extra dimensions) and $D \geqslant 3$ is different.
It was already
known that for the $\Lambda$-term case there is a regime with ``stabilization'' of extra dimensions, where
the expansion rate of the three-dimensional subspace as well as the scale factor (the ``size'') associated with extra dimensions reach constant value.
This regime is achieved if the curvature of the extra dimensions is negative. We demonstrate that it take place only if the number of extra dimensions is $D \geqslant 3$.
In the second part of the paper we study the influence of initial anisotropy. Our study reveals that the transition from Gauss-Bonnet Kasner
regime to anisotropic exponential expansion (with expanding
three and contracting extra dimensions) is stable with respect to breaking the symmetry within both three- and extra-dimensional subspaces. However, the details of the dynamics in $D=2$ and $D \geqslant 3$
are different.
Combining the two described affects allows us to construct a scenario in $D \geqslant 3$, where isotropisation of outer and inner
subspaces is reached dynamically from rather general anisotropic initial conditions.
\end{abstract}

\pacs{04.50.Kd, 11.25.Mj, 98.80.Cq}






\maketitle

\section{Introduction}

Extra-dimensional theories have been known~\cite{Nord1914} even prior to the General Relativity (GR)~\cite{einst}, but relatively known they become after works by Kaluza and
Klein~\cite{KK1, KK2, KK3}. Since then the extra-dimensional theories evolve a lot but the main motivation behind them remains the same -- unification of interactions. Nowadays one of the
promising candidate for unified theory is M/string theory.

Presence of the curvature-squared corrections in the Lagrangian of the gravitational counterpart of string theories is one of their distinguishing features.
Scherk and Schwarz~\cite{sch-sch} demonstrated the need for the $R^2$ and
$R_{\mu \nu} R^{\mu \nu}$ terms, while later Candelas et al.~\cite{Candelas_etal} proved the same for $R^{\mu \nu \lambda \rho} R_{\mu \nu \lambda \rho}$. Later it was
demonstrated~\cite{zwiebach} that the only
combination of quadratic terms that leads to a ghost-free nontrivial gravitation
interaction is the Gauss-Bonnet (GB) term:

$$
L_{GB} = L_2 = R_{\mu \nu \lambda \rho} R^{\mu \nu \lambda \rho} - 4 R_{\mu \nu} R^{\mu \nu} + R^2.
$$

\noindent This term, first found
by Lanczos~\cite{Lanczos1, Lanczos2} (therefore it is sometimes referred to
as the Lanczos term) is an Euler topological invariant in (3+1)-dimensional
space-time, but not in (4+1) and higher dimensions.
Zumino~\cite{zumino} extended Zwiebach's result on
higher-than-squared curvature terms, supporting the idea that the low-energy limit of the unified
theory might have a Lagrangian density as a sum of contributions of different powers of curvature. In this regard the Einstein-Gauss-Bonnet (EGB) gravity could be seen as a subcase of more general Lovelock
gravity~\cite{Lovelock}, but in the current paper we restrain ourselves with only quadratic corrections and so to the EGB case.

While considering extra-dimensional theories, regardless of the model, we need to explain where are additional dimensions. Indeed, with our current level of experiments, we clearly sense three
spatial dimensions and sense no presence of extra dimensions. The common explanation is that they are ``compactified'', meaning that they are so small that we cannot detect them. Perhaps, the
simplest class of such theories are the theories with ``spontaneous compactification''. Exact solutions of this class have been known for a long time~\cite{add_1}, but especially relevant for
cosmology are those with dynamical size of extra dimensions (see~\cite{add_4, Deruelle2, add_10, add_8} for different models). Notable recent studies include~\cite{add13}, where dynamical
compactification of the (5+1) Einstein-Gauss-Bonnet model was considered, \cite{MO04, MO14}, where different metric {\it Ans\"atze} for scale factors
corresponding to (3+1)- and extra-dimensional parts were studied and \cite{CGP1, CGP2, CGPT}, where we investigated general (e.g., without any {\it Ansatz}) scale factors and curved manifolds.
Also, apart from
cosmology, the recent analysis has focused on
properties of black holes in Gauss-Bonnet~\cite{alpha_12, add_rec_1, add_rec_2, addn_1, addn_2} and Lovelock~\cite{add_rec_3, add_rec_4, addn_3, addn_4, addn_4.1} gravities, features of
gravitational collapse in these theories~\cite{addn_5, addn_6, addn_7}, general features of spherical-symmetric solutions~\cite{addn_8}, and many others.

When it comes to exact cosmological solutions, two most common {\it Ansatz} used for the scale factor are exponential and power law. Exponential solutions represent de Sitter asymptotic stages
while power-law -- Friedmann-like.
Power-law solutions have been analyzed  in \cite{Deruelle1, Deruelle2} and more  recently in~\cite{mpla09, prd09, Ivashchuk, prd10, grg10} so that by now there is an  almost complete
description of the solutions of this kind (see also~\cite{PT} for comments regarding physical branches of the power-law solutions). One of the first considerations of the extra-dimensional
exponential solutions was done by Ishihara~\cite{Is86}; later considerations include~\cite{KPT}, as well as the models with both variable~\cite{CPT1} and constant~\cite{CST2} volume;
the general scheme for
constructing solutions in EGB gravity was developed and generalized for general Lovelock gravity of any order and in any dimensions~\cite{CPT3}. Also, the stability of the solutions was
addressed in~\cite{my15}
(see also~\cite{iv16} for stability of general exponential
solutions in EGB gravity), and it was
demonstrated that only a handful of the solutions could be called ``stable'', while the most of them are either unstable or have neutral/marginal stability.

If we want to find all possible regimes in EGB cosmology, we need to go beyond an exponential or power-law {\it Ansatz} and keep the scale factor generic.
We are particularly interested in models that allow dynamical compactification, so that we
consider the spatial part as the warped product of a three-dimensional and extra-dimensional parts. In that case the three-dimensional part is ``our Universe'' and we expect for this part to
expand
while the extra-dimensional part should be suppressed in size with respect to the three-dimensional one. In~\cite{CGP1} we demonstrated the there existence of regime when the curvature of the
extra dimensions is negative and the Einstein-Gauss-Bonnet theory does not admit a maximally symmetric solution. In this case both the
three-dimensional Hubble parameter and the extra-dimensional scale factor asymptotically tend to the constant values. In~\cite{CGP2} we performed a detailed analysis of the cosmological
dynamics in this model
with generic couplings. Later in~\cite{CGPT} we studied this model and demonstrated that, with an additional constraint on couplings, Friedmann late-time dynamics in three-dimensional part
could be restored.

Recently we have performed full-scale investigation of the spatially-flat cosmological models in EGB gravity with the spatial part being warped product of a three-dimensional and
extra-dimensional parts~\cite{my16a, my16b, my17a}. In~\cite{my16a} we demonstrated that the vacuum model has two physically viable regimes -- first of them is the smooth transition from
high-energy GB Kasner to low-energy GR Kasner. This regime
appears for $\alpha > 0$ at $D=1,\,2$ (the number of extra dimensions) and for $\alpha < 0$ at $D \geqslant 2$ (so that at $D=2$ it appears for both signs of $\alpha$). The other viable regime
is smooth transition from high-energy GB
Kasner to anisotropic exponential regime with expanding three-dimensional section (``our Universe'') and contracting extra dimensions; this regime occurs only for $\alpha > 0$ and at
$D \geqslant 2$. In~\cite{my16b, my17a} we considered $\Lambda$-term case and it appears that only realistic regime is the transition from high-energy GB Kasner to anisotropic exponential
regime; the low-energy GR Kasner is
forbidden in the presence of the $\Lambda$-term so the corresponding transition do not occur. Also, if we consider joint constraints on $(\alpha, \Lambda)$ from our cosmological analysis and
a black holes properties, different aspects of AdS/CFT and related theories in the presence of Gauss-Bonnet term (see~\cite{alpha_01, alpha_02, alpha_03, alpha_04, alpha_05, alpha_06, alpha_07,
alpha_08, add_rec_2, add_rec_4, dS}), the resulting bounds on $(\alpha, \Lambda)$ are (see~\cite{my17a} for details)

\begin{equation}
\begin{array}{l}
\alpha > 0, \quad D \geqslant 2, \quad \dac{3D^2 - 7D + 6}{4D(D-1)}  \equiv \eta_0 \geqslant \alpha\Lambda \geqslant \eta_2 \equiv - \dac{(D+2)(D+3)(D^2 + 5D + 12)}{8(D^2 + 3D + 6)^2},
\end{array} \label{alpha_limit3}
\end{equation}

\noindent where $\alpha$ is the Gauss-Bonnet coupling and $D$ is the number of extra dimensions.

The current paper is a natural continuation of our previous research on the properties of cosmological dynamics in EGB gravity. After a thorough investigation of spatially-flat cases
in~\cite{my16a, my16b, my17a}, it is natural to consider spatially non-flat cases.
Indeed, the spatial curvature affects inflation~\cite{infl1, infl2}, so that
it could change asymptotic regimes in other high-energy stages of the
Universe evolution, and we are considering one of them.
We already investigated the cases with negative curvature of the extra dimensions in~\cite{CGP1, CGP2, CGPT}, but
to complete description it is necessary to consider all possible cases.  We are going to consider all possible  curvature combination to see their influence on the dynamics -- we know the
regime for the case with both subspaces being spatially flat and will see the change in the dynamics with the curvatures being non-flat. This allows us to find all possible asymptotic regimes
in spatially non-flat case; together with the results for the flat case, it will complete this topic.

Another important issue we are going to consider is the anisotropy within subspaces. Indeed, the analysis in~\cite{my16a, my16b, my17a} is performed under conjecture that both three- and
extra-dimensional subspaces are isotropic. The question is, if the results are stable under small (or not very small) deviations of isotropy of these subspaces.
Finally, if we consider both effects, we could build two-steps scheme which allows us to qualitatively describe the dynamical compactification of anisotropic curved space-time.

The structure of the manuscript is as follows: first we write down the equations of motion for the case under consideration. Next, we study the effects of curvature -- we add all possible
curvature combinations to all known existing flat regimes and describe the changes in the dynamics. After that we draw conclusions for separately vacuum and $\Lambda$-term regimes and
describe their differences and generalities. After that we investigate the effects of anisotropy and find stability areas for different cases. Finally, we use both
effects to build two-steps scheme which allow us to describe the dynamics of a wide class spatially curved models.
In the end, we discuss the results obtained and draw the conclusions.

\section{Equations of motion}
\label{s2}

Lovelock gravity~\cite{Lovelock} has the following structure: its Lagrangian is constructed from terms

\begin{equation}
L_n = \frac{1}{2^n}\delta^{i_1 i_2 \dots i_{2n}}_{j_1 j_2 \dots
j_{2n}} R^{j_1 j_2}_{i_1 i_2}
 \dots R^{j_{2n-1} j_{2n}}_{i_{2n-1} i_{2n}}, \label{lov_lagr}
\end{equation}

\noindent where $\delta^{i_1 i_2 \dots i_{2n}}_{j_1 j_2 \dots
j_{2n}}$ is the generalized Kronecker delta of the order $2n$.
One can verify that $L_n$ is Euler invariant in $D < 2n$ spatial dimensions and so it would not give nontrivial contribution into the equations of motion. So that the
Lagrangian density for any given $D$ spatial dimensions is sum of all Lovelock invariants (\ref{lov_lagr}) upto $n=\[\dac{D}{2}\]$ which give nontrivial contributions
into equations of motion:

\begin{equation}
{\cal L}= \sqrt{-g} \sum_n c_n L_n, \label{lagr}
\end{equation}

\noindent where $g$ is the determinant of metric tensor,
$c_n$ is a coupling constant of the order of Planck length in $2n$
dimensions and summation over all $n$ in consideration is assumed.

The {\it ansatz} for the metric is
\begin{equation}
ds^{2}=-dt^{2}+a(t)^{2}d\Sigma_{(3)}^{2}+b(t)^{2}d\Sigma_{(\mathbf{D}%
)}^{2}\ ,  \label{Ansatz-metric}
\end{equation}%

\noindent where $d\Sigma _{(3)}^{2}$ and $d\Sigma _{(\mathbf{D})}^{2}$ stand for the
metric of two constant curvature manifolds $\Sigma _{(3)}$ and $\Sigma_{(\mathbf{D})}$\footnote{We consider {\it ansatz} for space-time in form of a warped product
\mbox{$M_4\times b(t)M_D$}, where $M_4$ is a Friedmann-Robertson-Walker manifold with scale factor $a(t)$ whereas
$M_D$ is a $D$-dimensional Euclidean compact and constant curvature manifold with scale factor $b(t)$.}. It is worth to point out that even a negative constant curvature space
can be
compactified by making the quotient of the space by a
freely acting discrete subgroup of $O(D,1)$ \cite{wolf}.

The complete derivation of the equations of motion could be found in our previous papers, dedicated to the description of the particular regime which appears in this
model~\cite{CGP1, CGP2}. It is convenient to use the following notation

\begin{eqnarray}
A_{(1)} &=&\frac{\overset{..}{a}}{a},\ \ \ C=\frac{\overset{.}{a}\overset{.}{%
b}}{ab},\ \ \ B_{(1)}=\frac{\overset{..}{b}}{b}, \notag \\
A_{(2)} &=&\frac{\left[ \gamma _{\left( 3\right) }+\left( \overset{.}{a}%
\right) ^{2}\right] }{a^{2}},\
\ \ B_{(2)}=\frac{\left[ \gamma _{\left( \mathbf{D}\right) }+\left( \overset{.}{b}%
\right) ^{2}\right] }{b^{2}}\label{ABdef}
\end{eqnarray}

\noindent and the following rescaling of the coupling constants

\begin{equation}
\alpha =\frac{\left(D+3\right) \left(D+2\right) \left(D+1\right) }{6}c_{0}\ ,\ \ \ \beta =\frac{\left(D+1\right) D\left(D-1\right) }{6}c_{1}\ ,\ \ \ \gamma =\frac{\left(D-1\right) \left(D-2\right)
\left(D-3\right) }{6}c_{2}\ . \label{abc}
\end{equation}%

Then, the equations of motion could be written in the following form:

\begin{equation}
\begin{array}{l}
\mathcal{E}_{0} =0\Leftrightarrow 0=\alpha +\beta \left( B_{(2)}+\dac{6}{%
D-1}C+\dac{6}{D\left(D-1\right) }A_{(2)}\right)
+\gamma \left(  B_{(2)}^{2}+\dac{12A_{(2)}B_{(2)}}{\left(
D-2\right) \left(D-3\right) }+\right.  \\ \\
\left. +\dac{24C^{2}}{\left(D-2\right) \left(D
-3\right) }+\dac{12B_{(2)}C}{\left(D-3\right) }+\dac{24A_{(2)}C}{%
\left(D-1\right) \left(D-2\right) \left(D
-3\right) }\right) ,  \label{eq0}
\end{array}
\end{equation}
\begin{equation*}
\mathcal{E}_{i}=0\Leftrightarrow 0=\alpha +\beta \left( B_{(2)}+\frac{
4A_{(1)}}{D\left(D-1\right) }+\frac{2B_{(1)}}{D-1
}+\frac{2A_{(2)}}{D\left(D-1\right) }+\frac{4C}{\left(
D-1\right) }\right) +\gamma \left(  B_{(2)}^{2}+\right.
\end{equation*}%
\begin{equation*}
+\frac{16A_{(1)}C}{\left(D-1\right) \left(D-2\right)
\left(D-3\right) }+\frac{8B_{(2)}C}{D-3}++\frac{
8A_{(1)}B_{(2)}}{\left(D-2\right) \left(D-3\right) }+
\frac{8A_{(2)}B_{(1)}}{\left(D-1\right) \left(D-2\right)
\left(D-3\right) }+
\end{equation*}%
\begin{equation}
\left. +\frac{16B_{(1)}C}{\left(D-2\right) \left(D
-3\right) }+\frac{4B_{(1)}B_{(2)}}{\left(D-3\right) }+\frac{
4A_{(2)}B_{(2)}}{\left(D-2\right) \left(D-3\right) }+
\frac{8C^{2}}{\left(D-2\right) \left(D-3\right) }\right)
\ \ ,  \label{eqq1}
\end{equation}%
while the equation $\mathcal{E}_{a}=0$ reads%
\begin{equation*}
\mathcal{E}_{a}=0\Leftrightarrow 0=\frac{D}{\left(D-
4\right) }\alpha +\frac{\left(D-2\right) }{\left(D-
4\right) }\beta \left( B_{(2)}+\frac{6A_{(1)}}{\left(D-1\right)
\left(D-2\right) }+\frac{2B_{(1)}}{D-2}+\frac{6A_{(2)}}{%
\left(D-1\right) \left(D-2\right) }+\frac{6C}{\left(
D-2\right) }\right) +
\end{equation*}%
\begin{equation*}
+\gamma \left( B_{(2)}^{2}+\frac{48A_{(1)}C}{\left(D
-2\right) \left(D-3\right) \left(D-4\right) }+\frac{
12B_{(2)}C}{D-4}+\frac{24C^{2}}{\left(D-3\right) \left(
D-4\right) }+\right.
\end{equation*}%
\begin{equation*}
+\frac{12A_{(1)}B_{(2)}}{\left(D-3\right) \left(D
-4\right) }+\frac{24A_{(2)}B_{(1)}}{\left(D-2\right) \left(
D-3\right) \left(D-4\right) }+\frac{24B_{(1)}C}{\left(
D-3\right) \left(D-4\right) }+\frac{4B_{(1)}B_{(2)}}{%
\left(D-4\right) }+
\end{equation*}%
\begin{equation}
\left. +\frac{12A_{(2)}B_{(2)}}{\left(D-3\right) \left(D
-4\right) }+\frac{24A_{(2)}C}{\left(D-2\right) \left(D-
3\right) \left(D-4\right) }+\frac{24A_{(1)}A_{(2)}}{\left(
D-1\right) \left(D-2\right) \left(D-3\right) \left(
D-4\right) }\right).   \label{eqq2}
\end{equation}

\section{Influence of curvature}
\label{s3}

In this section we investigate the impact of the spatial curvature on the cosmological regimes. As a ``background'' we use the results obtained in~\cite{my16a, my16b, my17a} -- exact
regimes for $\gammat = \gammad \equiv 0$ for both vacuum and $\Lambda$-term cases. As we use them as a ``background'' solutions, it is worth to quickly describe them all. All solutions found for both vacuum and $\Lambda$-term
cases could be splitted into two groups -- those with ``standard'' regimes as both past and future asymptotes and those with nonstandard singularity as one (or both) of the asymptotes. By the ``standard'' regimes we mean
Kasner (generalized power-law) and exponential. In our study me encounter two different Kasner regimes -- ``classical'' GR Kasner regime (with $\sum p_i = \sum p_i^2 = 1$ where $p_i$ is Kasner exponent from the definition
of power-law behavior $a_i (t) = t^{p_i}$), which we denote as $K_1$ (as $\sum p_i = 1) $ and it is low-energy regime; and GB Kasner regime (with $\sum p_i = 3$), which we denote as $K_3$ and it is high-energy regime. For realistic cosmology we should have high-energy regime as past asymptote and low-energy as future, but our investigation demonstrates that potentially both $K_1$ and $K_3$ could play a role as past and future asymptotes~\cite{my16a}. Also we should note that $K_1$ exist only in the vacuum regime, while $K_3$ as past asymptotes we encounter in both vacuum and $\Lambda$-term regimes (see~\cite{my16b} for details). The exponential regimes (where scale factors depend upon time exponentially, so Hubble parameters are constant) could be seen in both vacuum and $\Lambda$-term regimes and there are two of them -- isotropic and anisotropic ones. The former of them corresponds to the case where all the directions are isotropized and, since we
work in the multidimensional case, it does not fit the observations. On contrary, the latter of them have different Hubble parameters for three- and extra-dimensional subspaces. For realistic compactification we demand
expansion of the three- and contraction of the extra-dimensional spaces. The exponential solutions are denoted as $E_{iso}$ for isotropic and $E_{3+D}$ for anisotropic, where $D$ is the number of extra dimensions (so that, say, in $D=2$ the anisotropic exponential solution is denoted as $E_{3+2}$).

The second large group are the regimes which have nonstandard singularity as either of the asymptotes or even both of them. The nonstandard singularity is the situation which arises in nonlinear theories and in our particular
case it corresponds to the point of the evolution where $\dot H$ (the derivative of the Hubble parameter) diverges at the final $H$; we denote it as $nS$. This kind of singularity is ``weak'' by Tipler's classification~\cite{Tipler} and is
type II in classification by Kitaura and Wheeler~\cite{KW1, KW2}. Our previous research reveals that nonstandard singularity is a wide-spread phenomena in EGB cosmology, for instance, in $(4+1)$-dimensional Bianchi-I vacuum
case all the trajectories have $nS$ as either past or future asymptote~\cite{prd10}. Since a nonstandard
singularity means the beginning or end of dynamical evolution, either higher or lower values
of $H$  do not reached and so the entire evolution from high to low
energies cannot be restored; for this reason we disregard the trajectories with $nS$ in the present paper.

So that the viable (or realistic) regimes are limited to $K_3 \to K_1$ and $K_3 \to E_{3+D}$ for vacuum case and $K_3 \to E_{3+D}$ for $\Lambda$-term; these regimes we further investigate in the presence of curvature.

\subsection{Vacuum $K_3 \to K_1$ transition with curvature}

First we want to investigate the influence of the curvature on the vacuum Kasner transition -- transition from Gauss-Bonnet Kasner regime $K_3$ to standard GR Kasner $K_1$. We add curvature to either and both three- and
extra-dimensional manifolds and see the changes in the regimes. We label the cases as $(\gamma_3, \gamma_D)$ where $\gamma_3$ is the spatial curvature of the three-dimensional manifold and $\gamma_D$ -- of the extra-dimensional.
So that for $(0, 0)$ -- flat case -- we have $K_3 \to K_1$, as reported in~\cite{my16a}. Now if we introduce nonzero curvature, both $(1, 0)$ and $(-1, 0)$ do not change the regime and it remains $K_3\to K_1$. So that we can conclude
that $\gamma_3$ alone do not affect the dynamics. On contrary, $\gamma_D$ does -- $(0, 1)$ has the transition changes to $K_3 \to K_3^S$ (finite-time future singularity of the power-law type with $K_3$ behavior -- analogue of the recollapse from the standard cosmology), while
$(0, -1)$ change the transition to $K_3 \to K_D$. This $K_D$ is a new but non-viable regime with $p_3 \to 0$ and $p_D \to 1$ -- regime with constant-size three dimensions and expanding as power-law extra dimensions, which makes 
the behavior in the expanding subspace Milne-like, caused by the negative curvature. So that
the curvature of the extra dimensions alone makes future asymptotes non-viable. If we include both curvatures, the situation changes as follows: for $(1, 1)$ we have $K_3 \to K_3^S$; for $(1, -1)$ it is
$K_3 \to K_D$; for $(-1, 1)$ it is $K_3 \to K_3^S$ and finally for $(-1, -1)$ it is $K_3 \to K_{D+3}^{iso}$.

The described regimes require some explanations. First of all, as we reported in~\cite{my16a}, viable regimes have $p_a > 0$ and $p_D < 0$ -- indeed, we want expanding three-dimensional space and contracting extra dimensions to
achieve compactification. Then, it is clear why $\gamma_3$ alone does not change anything -- with expanding scale factor, the effect of curvature vanishes. It is also clear why $\gamma_D = +1$ makes $K_3^S$ as future asymptote
-- positive spatial curvature prevents infinite contracting of the extra dimensions and gives rise to new regime. But the most interesting is the effect of $\gamma_D = -1$ -- indeed, negative curvature not just stops the
contraction of the extra dimensions but starts their expansion, which change the entire dynamics drastically. Now extra-dimensional scale factor ``dominates'' and three-dimensional goes for a constant. It is like that for
zeroth and positive curvatures of the three-dimensional subspace, but for $\gamma_3 = -1$ -- so if both subspaces have negative curvature -- three-dimensional scale factor also start to expand due to the negative curvature,
leading to isotropic power-law solution $K_{D+3}^{iso}$, caused by the negative curvature.

The scheme above has one interesting feature -- as we described, $\gammad < 0$ give rise to regime with $p_3 \to 0$ and $p_D \to 1$ -- but in $D=3$ this gives us ``would be'' viable regime -- indeed, if both subspaces
are three-dimensional, as long as one is expanding and another is not, we could just call expanding one as ``our Universe'' and stabilized -- ``extra dimensions''. So that in $D=3$ there exist a regime with stabilized extra
dimensions and power-law expanding three-dimensional ``our Universe''. However, viability of this regime needs
more checks, and we leave this question to further study.

So that negative curvature of the extra dimensions gives rise to two new and interesting regimes -- $K_D$ with expanding extra dimensions and constant-sized three-dimensional subspace, and $K_{D+3}^{iso}$ -- isotropic
power-law solution. Both of them are not presented in the spatially-flat vacuum case, but also both of them are non-viable, so that they do not improve the chances for successful compactification. The only viable case is $K_3 \to K_1$ which
remains unchanged for $\gamma_D = 0$.

\subsection{Vacuum $K_3 \to E_{3+D}$ transition with curvature}

Now let us examine the effect of curvature on another viable vacuum regime -- transition from GB Kasner $K_3$ to anisotropic exponential solution $E_{3+D}$. Similar to the previously considered cases, for an anisotropic
exponential solution to be considered as ``viable'', we demand the expansion rate of the three-dimensional subspace to be positive while for extra dimensions -- to be negative. Let us see what  happens if we add nonzero
spatial curvature.

Similar to the previous case, the curvature of the three-dimensional subspace $\gamma_3$ alone does not change the dynamics -- $(1, 0)$ and $(-1, 0)$ both have $K_3 \to E_{3+D}$ regime. But unlike the previous case,
the curvature of the extra dimensions $\gamma_D$ alone makes the future asymptotes singular -- power-law-type finite-time future singularity in case of $\gamma_D = +1$ and nonstandard singularity in case of $\gamma_D = -1$.
The same situation remains in cases with both subspaces have curvature -- as long as $\gamma_D \ne 0$, the future asymptote is singular -- either power-law or nonstandard, depending on the sign of the curvature.

So that, similar to the previous case, the only viable regime is unchanged $K_3 \to E_{3+D}$ which occurs if $\gamma_D = 0$. But unlike previous case, this one does not give us interesting nonsingular regimes.

\subsection{$\Lambda$-term $K_3 \to E_{3+D}$ transition with curvature}

Finally, let us describe the effect of curvature on the only viable $\Lambda$-term regime -- $K_3 \to E_{3+D}$ transition described in~\cite{my16b, my17a}. The condition for viability is the same as in the described above cases --
expansion of the three-dimensional subspace and contraction of the extra dimensions. Our investigation suggests that the cases with $D=2$ and $D\geqslant 3$ are different; let us first describe $D=2$ case.
According to~\cite{my16b, my17a}, there are three domains for the $\Lambda$-term case where $K_3 \to E_{3+D}$ transition take place -- i) $\alpha > 0$, $\Lambda > 0$, $\alpha \Lambda \leqslant \zeta_0$ with $\zeta_0 = 1/2$ for
$D=2, 3$ and $\zeta_0 = (3D^2 - 7D + 6)/(4D(D-1))$, ii) entire $\alpha > 0$, $\Lambda < 0$ domain and iii) $\alpha < 0$, $\Lambda > 0$, $\alpha \Lambda \leqslant -3/2$. Formally i) and ii) supplement each other to form a single
domain $\alpha > 0$, $\alpha \Lambda \leqslant \zeta_0$, but in i) there also exist isotropic exponential solutions, which, as we will see, affects the dynamics, so we consider these two domains separately. So for i) domain,
we have regime unchanged if $\gammad = 0$, isotropisation ($K_3 \to E_{iso}$) if $\gammad < 0$ and nonstandard singularity $nS$ if $\gammad > 0$. In ii) domain, we again have unchanged $K_3 \to E_{3+2}$ if $\gammad = 0$ and
$nS$ in all other (i.e. $\gammad \ne 0$) cases. Already here we can see the difference between i) and ii) domains. Finally, iii) domain have the same dynamics as ii). So that the domain where isotropic and anisotropic exponential
solutions coexist, we have slightly richer dynamics, but neither of the regimes are viable; the only viable regime is unchanged $K_3 \to E_{3+2}$ and it take place if $\gammad = 0$. Now if we consider
 general $D\geqslant 3$ case, the resulting regimes are as follows: now i) and ii) domains have the same structure -- opposite to the $D=2$ case, the structure is as follows -- the only viable regime is unchanged
 $K_3 \to E_{3+D}$ which exist if $\gammad = 0$; if $\gammad \ne 0$, we always have $nS$. The iii) domain have the structure: unchanged $K_3 \to E_{3+D}$ if $\gammad = 0$, ``stabilization'' (or ``geometric frustration''
 regime~\cite{CGP1, CGP2}) if $\gammad < 0$ and $nS$ if $\gammad > 0$. This ``stabilization'' regime
is the regime which naturally appears in the ``geometric frustration'' case and described in~\cite{CGP1, CGP2}. In this regime the Hubble parameter, associated with three-dimensional subspace, reach constant value while the Hubble parameter, associated with extra dimensions, reach zero (and so the corresponding scale factor -- the ``size'' of extra dimensions -- reach constant value; the size of extra dimensions ``stabilize'').

So that in this last case -- $\Lambda$-term $K_3 \to E_{3+D}$ transition -- the ``original'' regime remains unchanged for $\gammad = 0$. For nonzero curvature of extra dimensions, if it is positive, the future asymptote is singular, if it is negative, and $D \geqslant 3$, in future we could have the regime with stabilization of extra dimensions, otherwise it is also singular.

We remind a reader that the geometric frustration proposal suggests that the dynamical compactification
with stabilization of extra dimensions occurs only for those coupling constant in EGB gravity for
which maximally-symmetric solutions are absent.
In turn, absence of the maximally-symmetric
solutions means absence of the isotropic exponential solutions, so that with negative curvature of the extra dimensions, isotropic and anisotropic exponential solutions cannot ``coexist'', which
means that for any set of couplings and parameters, only one of them could exist.
 The validity of this proposal have been checked numerically in~\cite{my16b, my17a}
for larger number of extra dimensions, now we see that it is valid also for the $D=3$ case.

It is not the same in the flat case -- for instance, for $\alpha > 0$, $\Lambda > 0$~\cite{my16b, my17a} we have both $K_3 \to E_{iso}$
and $K_3 \to E_{3+D}$ on different branches. If we turn on the negative curvature $\gammad < 0$, the former of them remains while the latter turns to $K_3 \to nS$, nonstandard singularity in $D=2$, or to stabilization
regime in $D > 2$. This way we can see that $D=2$ is somehow pathological -- in presence of curvature, there are no realistic regimes in $D=2$ but there are in $D \geqslant 3$.

Finally, we made the same analysis starting from the exponential regime instead of the GB Kasner with the
same number of expanding and contracting dimensions. The final fate of all trajectories appears to be the
same. We will use this note later in the Sec.~\ref{s5}.

\subsection{Summary}

So that all three considered cases have the original regimes unchanged as long as $\gammad = 0$. This means that the curvature of the three-dimensional world alone cannot change the future asymptote. For nonzero curvature of the
extra dimensions, the situation is different in all three cases: in vacuum $K_3 \to E_{3+D}$ case all trajectories with $\gammad \ne 0$ are singular; in vacuum $K_3 \to K_1$ we have two new regimes but both of them
are non-viable; finally, in $\Lambda$-term $K_3 \to E_{3+D}$ case if $\gammad > 0$ the future asymptote is singular while for $\gammad < 0$ there could be viable regime with stabilization of extra
dimensions, but this regime occurs only when isotropic exponential solution cannot exist and in $D \geqslant 3$.

To conclude, it seems that the only important player in this case is the curvature of extra dimensions. And this is clear why is it so -- from requirements of viability we demand that three-dimensional subspace
should expand while extra dimensions should contract. The expansion of the three dimensions cannot be stopped neither by $\gammat > 0$ nor by $\gammat < 0$, that is why $\gammat$ does not influence on the dynamics. On the other hand, extra dimensions are contracting, so both signs of extra-dimensional
curvature affect it -- positive usually leads to singularity (standard or not) while negative could turn it to expansion (what we see in $K_D$ and $K_{3+D}^{iso}$ regimes). The latter could even change the dynamics in
three-dimensional sector, what we also see in $K_{3+D}^{iso}$ regime.

\section{Influence of anisotropy}
\label{s4}

In this section we address the problem of anisotropy of each subspaces. In this case the equations of motion are different from (\ref{eq0})--(\ref{eqq2}); the metric {\it ansatz} has the form

\begin{equation}\label{metric}
g_{\mu\nu} = \diag\{ -1, a_1^2(t), a_2^2(t),\ldots, a_n^2(t)\};
\end{equation}

\noindent substituting it into the Lagrangian and following the derivation described in Section~\ref{s2} gives us the equations of motion:

\begin{equation}
\begin{array}{l}
2 \[ \sum\limits_{j\ne i} (\dot H_j + H_j^2)
+ \sum\limits_{\substack{\{ k > l\} \\ \ne i}} H_k H_l \] + 8\alpha \[ \sum\limits_{j\ne i} (\dot H_j + H_j^2) \sum\limits_{\substack{\{k>l\} \\ \ne \{i, j\}}} H_k H_l +
3 \sum\limits_{\substack{\{ k > l >  \\   m > n\} \ne i}} H_k H_l
H_m H_n \] - \Lambda = 0
\end{array} \label{dyn_gen}
\end{equation}

\noindent as the $i$th dynamical equation. The first Lovelock term---the Einstein-Hilbert contribution---is in the first set of brackets and the second term---Gauss-Bonnet---is in the second set;
$\alpha$
is the coupling constant for the Gauss-Bonnet contribution and we put the corresponding constant for Einstein-Hilbert contribution to unity.
Also, since in this section we consider spatially flat cosmological models, scale
factors do not hold much in the physical sense and the equations are rewritten in terms of the Hubble parameters $H_i = \dot a_i(t)/a_i(t)$. Apart from the dynamical equations, we write down the constraint equation

\begin{equation}
\begin{array}{l}
2 \sum\limits_{i > j} H_i H_j + 24\alpha \sum\limits_{i > j > k > l} H_i H_j H_k H_l = \Lambda.
\end{array} \label{con_gen}
\end{equation}

The relationship between ($c_0, c_1, c_2$) and ($\alpha, \Lambda$) is

\begin{equation}
\begin{array}{l}
c_0 = - \dac{6\Lambda}{(D+3)(D+2)(D+1)};\quad c_1 = \dac{6}{D+1};\quad c_2 = 6D\alpha.
\end{array} \label{rel}
\end{equation}

First, let us consider $D=2$ case -- it was demonstrated in~\cite{my16a, my16b, my17a} that $D=2$ case has all regimes
which higher-dimensional cases possess and does not have any extra regimes, so that $D=2$ case is the simplest representative case.
We seek an answer to the question -- if the subspaces are not exactly isotropic (we consider the spatial part being a
product of three- and two-dimensional isotropic subspaces), how it affect the dynamics? Is the asymptote is still reached or not? Indeed, totally anisotropic (Bianchi-I-type) cosmologies are more
generic, and if they still could lead to the asymptotes under consideration, this would wider the parameters and initial conditions spaces which could lead to viable compactification. Thorough
investigation of $D=1$ case revealed~\cite{prd10} that only $nS$ is available as a future asymptote in vacuum case (compare with~\cite{my16a} for regimes in $[3+1]$ spatial splitting), so that
the problem of ``loosing'' the regimes in case of broken symmetry exists.

To investigate this effect, we solve the general equations (i.e., without $H_1 = H_2 = H_3 = H$ and $H_5 = \cdots = H_{D-3} = h$ {\it ansatz} implied) in the vicinity
of the exact exponential and power-law solutions to see if the exact solution is reached in the course of the evolution, or if it is replaced with some other asymptote.

\begin{figure}
\includegraphics[width=1.0\textwidth, angle=0]{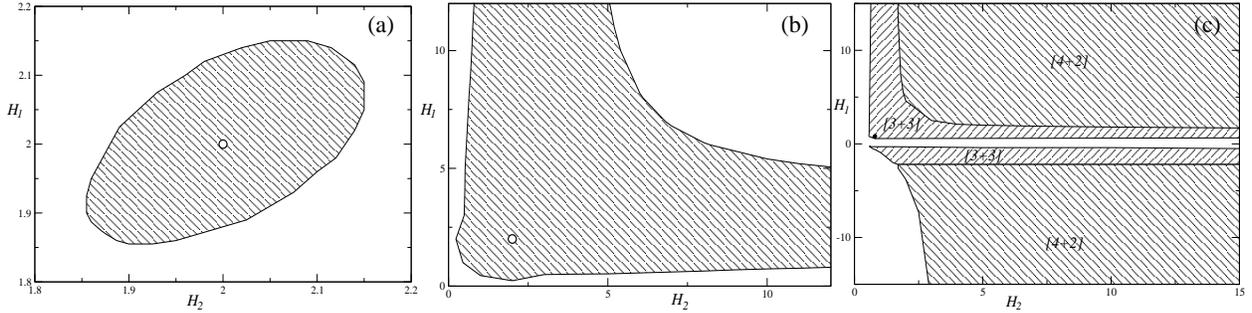}
\caption{Typical stability areas for vacuum $K_3\to K_1$ regime on (a) panel; vacuum and $\Lambda$-term $K_3\to E_{3+2}$ regime on (b) panel;
vacuum and $\Lambda$-term $K_3\to E_{3+3}$ (and possibly higher number of extra dimensions as well) regime on (c) panel
(see the text for more details).}\label{f2}
\end{figure}
\begin{figure}
\includegraphics[width=1.0\textwidth, angle=0]{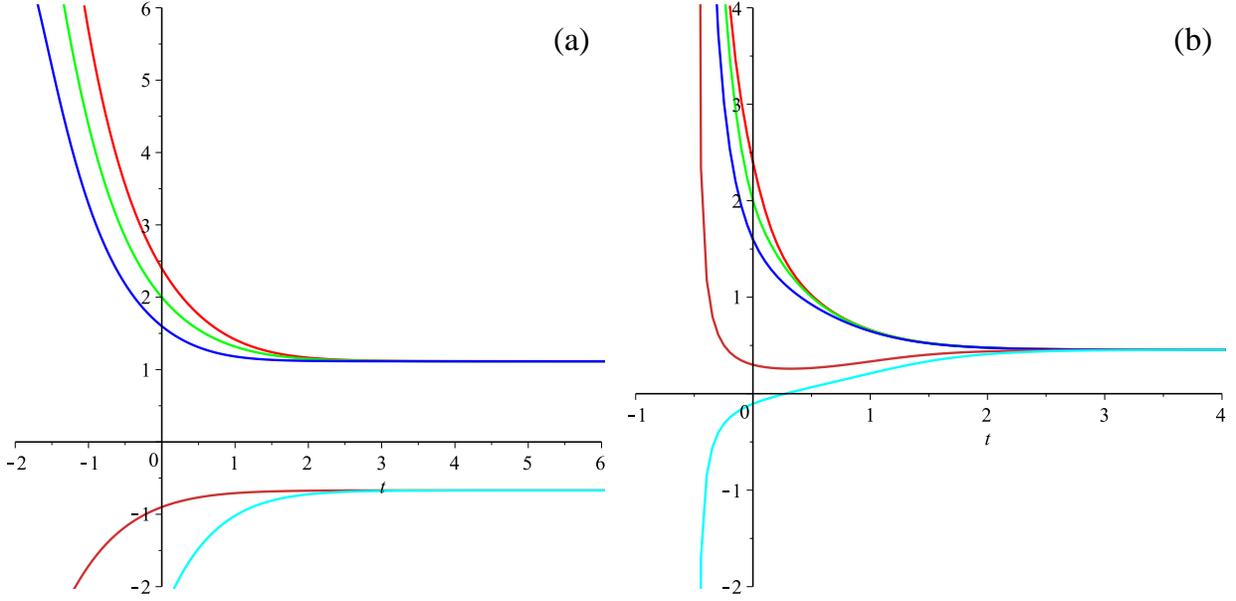}
\caption{Typical evolution curve for stable anisotropic (a) and isotropic (b) exponential solutions with broken symmetry in both subspaces in the $D=2$ case
(see the text for more details).}\label{f1}
\end{figure}

We start with vacuum regimes; according to~\cite{my16a}, in the vacuum $D=2$ case at high enough $H_0$ (initial value for the Hubble parameter, associated with three-dimensional
subspace), there are four combinations the branch (two of them, $h_1$ and $h_2$) and $\alpha \lessgtr 0$. First of the cases, $\alpha > 0$ and $h_1$, gives $K_3 \to K_1$
transition. If we
break the symmetries in both spaces, the stability of the regime is broken as well -- in Fig.~\ref{f2}(a) we presented the analysis of this case. There we present the regime depending on the
initial conditions -- we seek the regime change around $H_1 = H_2 = H_3 = 2.0$ exact solution and $H_4 = H_5 = h_0$ is being found from the constraint equation (\ref{eq0}); we fix $H_3 = 2.0$
and $H_4 = h_0$ and change $H_1$ and $H_2$ and find $H_5$ from constraint equation.
The exact solution in question ($H_1 = H_2 = H_3 = 2.0$, $H_4 = H_5 = h_0$) is depicted as a circle. The shaded area corresponds to $K_3\to K_1$ regime while the area which surrounds it --
to $K_3 \to nS$. One can see that the
stability region is quite small and any substantial deviation from the exact solution cause nonstandard singularity.
The second case, $\alpha > 0$ and $h_2$, have $K_3 \to E_{3+2}$ regime. With broken symmetry the regime is conserved much better then the previous one -- in Fig.~\ref{f2}(b) we presented the
analysis of this case. One can see that not just the area of the regime stability covers much large initial conditions, but this area is also unbounded. The typical evolution of such
transition is illustrated in Fig.~\ref{f1}(a).
The next case, $\alpha < 0$ and $h_1$, has $K_3 \to K_1$
transition, just like the first one, and their stability is similar.
Finally, the last case $\alpha < 0$ and $h_2$, governs $K_3 \to E_{iso}$ transition. If we break the symmetry for this case, the resulting stability area is quite similar to that of $K_3 \to K_1$.

To summarize the results for the vacuum case, only $K_3 \to E_{3+2}$ -- the transition from GB Kasner to anisotropic exponential solution -- is stable. All
other regimes -- transitions to isotropic exponential solution and to GR Kasner -- have much smaller stability areas and
could be called ``metastable''. Formally, the basin of attraction of $K_1$
and isotropic expansion is
nonzero and they are stable within it, but on the other hand its area is much smaller then
that of $E_{3+2}$; so that comparing with the two we decided to call $K_3
\to E_{3+2}$ as ``stable'' while $K_3 \to K_1$ and $K_3 \to E_{iso}$ as
``metastable''.


Now let us consider $\Lambda$-term case. According to~\cite{my16b}, in the presence of $\Lambda$-term the variety of the regimes is a bit different from the vacuum case. Again,
there are two branches ($h_1$ and $h_2$) and now in addition to variation in $\alpha$ there is variation in $\Lambda$ and in their product $\alpha\Lambda$.

The first case is $\alpha > 0$, $\Lambda > 0$. There on $h_1$ branch we have $K_3 \to E_{3+2}$ if $\alpha\Lambda \leqslant 1/2$ and $K_3 \to nS$ if $\alpha\Lambda > 1/2$. Another ($h_2$) branch
has $K_3 \to E_{iso}$ regardless of $\alpha\Lambda$. All these three branches are stable -- breaking the symmetry of both subspaces keeps the regimes as they are within wide vicinity of the
exact solution, like in Fig.~\ref{f2}(b).
Stable solution $K_3 \to E_{iso}$ as a future attractor for broken symmetry in both subspaces is illustrated in Fig.~\ref{f1}(b). The next case to consider is $\alpha > 0$, $\Lambda < 0$;
there $h_1$ branch has $K_3 \to E_{3+2}$
while $h_2$ has $K_3 \to nS$ and the former of them is proved to be stable (the latter is not viable so its stability is of little importance). Now let us turn to $\alpha < 0$ cases and the
first one is with $\Lambda > 0$. There at $\alpha\Lambda \geqslant -5/6$
both branches have $K_3 \to E_{iso}$ regime and both of them are metastable -- only the initial conditions which are very close to the exact solution lead to $E_{iso}$, those beyond lead to $nS$.
On contrary, at $\alpha\Lambda < -5/6$
on $h_1$ branch we have $K_3 \to E_{3+2}$ while on $h_2$ branch $K_3 \to nS$ and again $E_{3+2}$ is stable. Finally, $\alpha < 0$, $\Lambda < 0$ has $K_3 \to E_{iso}$ on $h_1$ and $K_3 \to nS$
on $h_2$ and in this case $E_{iso}$ is stable.

\begin{figure}
\includegraphics[width=1.0\textwidth, angle=0]{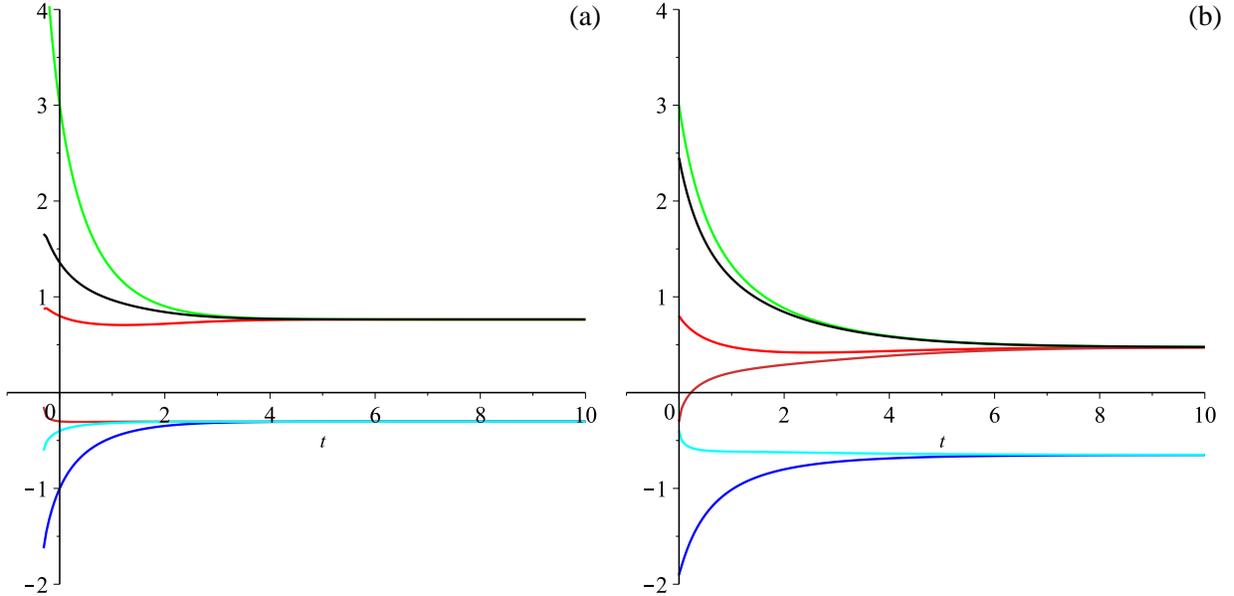}
\caption{Typical evolution curve for stable anisotropic $[3+3]$ (a) and $[4+2]$ (b) exponential solutions with broken symmetry in both subspaces in the $D=3$ case
(see the text for more details).}\label{f4}
\end{figure}

In addition to the described above $D=2$ case, we also considered $D=3$. The methodology is the same and the results for vacuum $K_3 \to K_1$ are also the same. But the results for both vacuum and $\Lambda$-term
$K_3 \to E_{3+3}$ transition are different and presented in Fig. \ref{f2}(c), where the initial conditions leading to $E_{3+3}$ are
shaded with $[3+3]$ note on them. One can see that the stability area is unbounded, as it was in $D=2$ case, but there are differences as well. First, the upper part seems shrinked in comparison with $D=2$ - so that starting from
a vicinity of the exact solution, it is less probable to end up on $E_{3+3}$. Instead, we have $K_3 \to E_{4+2}$ -- the exponential solution with four expanding and two contracting dimensions, which is, obviously, non-viable.
 In Fig.~\ref{f4} we presented $K_3 \to E_{3+3}$ in (a) panel and $K_3 \to E_{4+2}$ in (b) panel with the latter originates from some vast vicinity of the former.
We also have some initial conditions starting from the negative values to lead to the exponential solution (which could ``compensate'' the loss in the upper part) -- something we have never seen in $D=2$ -- but this is the effect of the number of dimensions
-- in $D=2$, due to the lesser number of dimensions, the constraint is more tight while in $D \geqslant 3$ it is more relaxed. The presence of the $E_{4+2}$ is also the effect of the higher number of extra dimensions -- indeed,
as we demonstrated in~\cite{CPT1}, in five spatial dimensions there is only one stable anisotropic exponential solution -- $E_{3+2}$ (see~\cite{my15, iv16} for stability issues), while in six and higher there are more~\cite{CPT3} and there is a chance to end up on another exponential solution. As the number of exact solutions grow up with the number of dimensions, in higher dimensions it is probable to end up on another exponential solution, rather then
$E_{3+D}$.

The black circle in Fig.~\ref{f2}(c) corresponds to the exact $E_{3+3}$ solution and one can see that the initial conditions are aligned along $H_i^{(0)} \sim H_j^{(0)}$. The same could be seen from $D=2$ case as well
(see Fig.~\ref{f2}(b)). The reason for it is quite clear -- indeed, with appropriate $H_i^{(0)} = H_j^{(0)}$ the exact $E_{3+D}$ solution is achieved explicitly, so that it is natural for the initial conditions to tend to this
relation.

To conclude, we see that all $K_3 \to E_{3+D}$ regimes in $\Lambda$-term case are stable with respect to breaking the symmetry of both subspaces. On the other hand, another nonsingular regime,
$K_3 \to E_{iso}$  is stable only for ($\alpha > 0$, $\Lambda > 0$) and ($\alpha < 0$, $\Lambda < 0$). Finally, $K_3 \to K_1$ in vacuum is also stable, but its basin of attraction is quite small and any
substantial deviation from the exact solution destroys it.

\section{Two-step scheme for general spatially curved case}
\label{s5}

The results of two previous sections allow us to construct a scenario of compactification
which satisfy two important requirements:

\begin{itemize}
\item the evolution starts from a rather general anisotropic initial conditions,
\item the evolution ends in a state with three isotropic big expanding dimensions and stabilized isotropic extra
dimensions.
\end{itemize}

The first part of the scenario in question uses the results of Sec.~\ref{s4}. We have seen there that while starting
from a state in the dashed zone of Fig.~\ref{f2}(b),(c) the flat anisotropic Universe tends to the exponential
solution with three equal expanding dimensions. The initial conditions for such a behavior are not so
restricted. From the Fig.~\ref{f2}(b) we can see that initial state should already have three expanding and two
shrinking dimensions, however, since all Gauss-Bonnet Kasner solutions (as well as usual GR Kasner  solutions) should have et least one shrinking
dimension, this requirement does not constraint possible initial state very seriously -- in any cases we
should expect that contracting dimensions are present in the initial conditions. Within this situation
the dashed zone occupy rather big part of initial condition space of Fig.~\ref{f2}(b), and any solution from this
zone ends up in exponential solution of desired type.

In higher dimensions the situation is from one side worsening -- as it is seen from Fig.~\ref{f2}(c), in $D \geqslant 3$ there are more then one stable anisotropic exponential solution, so that starting from the vicinity of exact $E_{3+3}$
solution we could end up in $E_{4+2}$ solution, which does not has realistic compactification. However,
from the other side, initial conditions with 2 expanding and 4 contracting dimensions can end up in 3+3
exponential solution.

Suppose also, that a negative spatial curvature is small enough at the beginning and starts to be important
only after this transition to exponential solution (which is established in the present paper only for
a flat Universe) already occurred. This condition allows us to glue the second part of the scenario which
requires negative spatial curvature of the inner space. We have see in Sec.~\ref{s3} that exponential solution
turn to the solution with stabilized extra dimensions in this case. As a result of these two stages
a Universe starting from initially anisotropic both outer expanding three dimensional space and contracting
 inner space evolves naturally to the final stage with isotropic three big dimensions and
isotropic and stabilized inner dimensions. The only additional condition for this scenario
to realize (in addition to starting from the appropriate zone in the initial conditions space) is that spatial curvature should
become dynamically important only after the transition to exponential solution occurs. As we mentioned in Sec.~\ref{s3}, this part (and so the entire scheme as well) works only for $D \geqslant 3$.


\section{Discussions and conclusions}

Prior to this paper, we completed study of the most simple (but the most important as well) cases. The spatial part of these cases is the product of three- and extra-dimensional subspaces
which are spatially flat and isotropic~\cite{my16a, my16b, my17a}. So that the obvious next step is consideration of these subspaces being non-flat and anisotropic, and that is what we have
done in current paper. Non-flatness is addressed by assuming that both subspaces have constant curvature while anisotropy -- by breaking the symmetry between the spatial directions. The
results of the curvature study suggest that the only viable regimes are those from the flat case with $\gammad = 0$ requirement. Additionally, in the $\Lambda$-term case there is
``geometric frustration'' regime, described in~\cite{CGP1, CGP2} and further investigated in~\cite{CGPT} with $\gammad < 0$ requirement.

Our study reveals that there is a difference between the cases with $\gammad = 0$ and $\gammad < 0$: the former of them have only exponential solutions and the isotropic and anisotropic solutions coexist; the
latter have the regime with stabilization of the extra dimensions (instead of ``pure'' anisotropic exponential regime) and isotropic exponential regimes cannot coexist with regimes of stabilization  -- this difference was not noted before.
The curvature effects also differ in different $D$ -- in $D=2$ there is no stabilization of extra dimensions while in $D \geqslant 3$ there is.

In $D=3$ and $\gammad < 0$ there is also an interesting regime in the vacuum case -- the regime with stabilization of one and power-law expansion of another three-dimensional subspaces; viability of this regime
for some compactification scenario needs further investigations.

The results of anisotropy study reveal that the $K_3 \to E_{3+D}$ regime is always stable with respect to breaking the isotropy in both subspaces, meaning that within some vicinity
of exact $K_3 \to E_{3+D}$ transition, all initial conditions still lead to this regime (see Fig.~\ref{f1}(a)). Though, the area of the basin of attraction for this regime depends on the number of extra dimensions $D$ --
in $D=2$ it is quite vast (see Fig.~\ref{f2}(b)) and there are no other anisotropic exponential solutions, in $D=3$ (and higher number of extra dimensions) it seems smaller\footnote{To quantitatively address this question we
need to introduce appropriate measure and since the area is unbounded, it is not an easy task. Also, the answer will depends on the chosen measure, so we skip the quantitative analysis.}
and there are initial conditions in the vicinity of
$E_{3+D}$ which leads to other exponential solutions. In our particular example $D=3$, presented in Fig.~\ref{f2}(c), some of the initial conditions from the vicinity of $E_{3+3}$ end up in $E_{4+2}$ instead. We expect that
in higher number of extra dimensions the situation for $E_{3+D}$ would be more complicated and requires a special analysis.

Another viable regime, $K_3 \to K_1$ from the vacuum case, as well as other
non-viable regimes, are ``metastable'' -- formally they are stable, but
their basin of attraction is much smaller compared to that of $E_{3+D}$ (see Fig.~\ref{f2}(a)).

Our study clearly demonstrates that the dynamics of the non-flat cosmologies could be different from flat cases and even some new regimes could emerge. In this paper we covered only the simplest
case with constant-curvature subspaces leaving the most complicated cases aside -- we are going to investigate some of them deeper in the papers to follow.


Now with both effects -- the spatial curvature and anisotropy within both subspaces -- being described, let us combine them. In the totally anisotropic case,
as we demonstrated, wide area of the initial conditions  leads to anisotropic exponential solution (for the values of couplings and parameters when isotropic exponential solutions
do not exist). So that if we start from the some vicinity of the exact exponential
solution, and if the initial scale factors are large enough for the curvature effects to be small, we shall reach the anisotropic exponential solution with expanding three and contracting
extra dimensions. After that the curvature effects in the expanding subspace are nullified while in the contracting dimensions they are not. If it is vacuum case, as we shown earlier, as long as
$\gammad \ne 0$ we encountered nonstandard singularity, so that the vacuum case is pathological in this scenario. In the $\Lambda$-term case, as we reported earlier, for $\gammad = 0$ we recover
the same exponential regime, for $\gammad > 0$ the behavior is singular and only for $\gammad < 0$ we obtain $E_{C, 0}$ -- ``geometric frustration'' scenario~\cite{CGP1, CGP2} with stabilization of the extra dimensions.

So that we can see that the proposed two-steps scheme works only for the $\Lambda$-term case and only if $\gammad < 0$ -- in all other cases it either provides trivial regimes, or leads to
singular behavior. Also, there is a minor problem with the number of extra dimensions -- as we noted, the first stage of this scheme -- reaching the exponential asymptote from initial anisotropy -- best achieved in $D=2$
and the probability of reaching $E_{3+D}$ could decrease with growth of $D$. On the other hand, the second stage -- when the negative curvature changes the contracting exponential solution for the extra dimensions into
stabilization -- is not presented in $D=2$ and only manifest itself in $D \geqslant 3$. So that the described two-stages scheme works only in $D \geqslant 3$ and in this case the initial conditions for the first stage are already
not so wide, though a fine-tuning of initial conditions is not needed.

This finalize our paper. The presented analysis suggests that more in-depth investigation of both curvature and anisotropy effects are required -- we have investigated and described
the most simple but still very important cases -- constant-curvature and
flat anisotropic (Bianchi-I-type) geometries; in the papers to follow we are
going to consider more complicated topologies.

\begin{acknowledgments}
The work of A.T. is supported by RFBR grant 17-02-01008 and
by the Russian Government
Program of Competitive Growth of Kazan Federal University. Authors are grateful to Alex Giacomini (ICFM-UACh, Valdivia, Chile)
for discussions.

\end{acknowledgments}

\end{document}